# Cross talk by extensive domain wall motion in arrays of ferroelectric nanocapacitors

Yunseok Kim,[1, a)] Hee Han,[2, b)] Ionela Vrejoiu,[1] Woo Lee,[2] Dietrich Hesse,[1] and Marin Alexe[1]

[1]*Max Planck Institute of Microstructure Physics, D-06120 Halle (Saale), Germany*

[2]*Korea Research Institute of Standards and Science (KRISS), Daejeon 305-340, Korea*

[a)]E-mail: kimy4@ornl.gov, present address: The Center for Nanophase Materials Sciences, Oak Ridge National Laboratory, Oak Ridge, Tennessee 37831, United States.
[b)]E-mail: h2m2h00@kriss.re.kr

We report on extensive domain wall motion in ferroelectric nanocapacitor arrays investigated by piezoresponse force microscopy. Under a much longer or higher bias voltage pulse, compared to typical switching pulse conditions, domain walls start to propagate into the neighbouring capacitors initiating a significant cross-talk. The propagation paths and the propagated area into the neighbouring capacitors were always the same under repeated runs. The experimental and the simulated results show that the observed cross-talk is related to the capacitor parameters combined with local defects. The results can be helpful to test the reliability of nanoscale ferroelectric memory devices.

Ferroelectric materials are most promising candidates for non-volatile memory devices due to a number of obvious advantages such as high density and fast read/write speed.[1] In particular, the minimum bit size of ferroelectric materials is theoretically only few nanometers across, which is very attractive for ultra-high density memory applications.[2] In applications such as ferroelectric memory devices, ferroelectric domain switching including domain wall motion plays an important role because it is directly related to the data bit signal processing.

In order to achieve terabit memory density, the size of ferroelectric capacitors needs to be downscaled into the nanoscale range, arriving at nanocapacitors. Subsequently, the nanoscale properties of these nanocapacitors need to be studied. Piezoresponse force microscopy (PFM) is well-known as a unique tool for nanoscale studies of the switching properties.[3,4] However, nanoscale studies of domain wall motion based on PFM have been primarily focused on microcapacitors.[5,6,7] The microcapacitors in previous reports possess very large aspect ratios since, while the film has nanoscale thickness, the capacitor size is still in the micron-scale range. For typical microcapacitor size and switching pulse conditions, domain switching is always confined to the capacitor boundaries.[8] However, since both film thickness and capacitor size have to be decreased to achieve a high memory density, the domain wall motion of the nanocapacitors may be different from that of microcapacitors. This is mostly due to size effects associated with small capacitor dimensions, such as extended structural defects that would be able to more significantly influence the domain wall motion of the nanocapacitors. Indeed, we have recently observed that the domain switching dynamics of nanocapacitors did not follow the classical switching model which is valid on the micron-scale counterparts.[9] Therefore, a detailed investigation of the domain wall motion in arrays of nanocapacitors is required.

Here, we show extensive domain wall motion towards neighbouring capacitors within arrays of sub-100 nm ferroelectric capacitors resulting in significant cross-talk. The cross-talk phenomenon has an important impact on actual devices, since it can affect the accuracy of the recorded bit signals in neighboring capacitors.

35 nm thick epitaxial $Pb(Zr_{0.2}Ti_{0.8})O_3$ (PZT) and 75 nm thick epitaxial $BiFeO_3$ (BFO) thin films were deposited by pulsed laser deposition on epitaxially grown Pt (001)/MgO (001) and $SrRuO_3$ (001)/$DyScO_3$ (110) single-crystal substrates, respectively.[8,9,10] Nanocapacitors on both PZT and BFO thin films were fabricated by evaporating 15-25 nm thick Pt top electrodes at room temperature through an ultra-thin anodic aluminium oxide (AAO) stencil mask.[8,9,11] The pore sizes of the AAO masks were about 70 nm and 60 nm for the PZT and BFO nanocapacitors, respectively, and the inter-pore distances of the AAO masks were about 100 nm for both types of nanocapacitors. The aspect ratio, i.e. the ratio between the capacitor diameter and the film thickness, was 2 and 0.8 for PZT and BFO, respectively. Details of PFM measurements can be found elsewhere.[8-10] The finite element analysis was performed by commercial software (COMSOL Multiphysics, COMSOL). The boundary conditions of the model were the same as in Ref. 11.

Figure 1(a) shows instantaneous domain configurations of the PZT nanocapacitors for the background poling (*i*) and the subsequent domain switching (*ii-vi*). As shown in Fig. 1(a)(*ii*), the switched domain of the film-type nanocapacitor is initially confined by the capacitor boundary.[10,11] For an applied bias of -4 V, it turned out that a pulse width below 1 ms usually was sufficient to switch the entire area of the capacitor and the switched domain was always confined by the capacitor boundary. However, when a much longer or higher bias pulse compared to typical switching pulse conditions was applied to the target capacitor, the domain wall started to propagate into the neighbouring capacitors [Figs. 1(a)(*iii-vi*)] generating a significant cross-talk. Intriguingly, the propagation paths to the neighbouring capacitors and the switched portion inside each affected neighbouring capacitor were always the same under repeated runs. It was also observed for BFO nanocapacitors as presented in Fig. 1(b). Similarly, at a bias pulse with an amplitude higher than -4 V applied to the target capacitor, the domain wall started to propagate into the neighbouring capacitors. On increasing the bias pulse amplitude the number of the affected neighbouring capacitors increased. However, for the BFO nanocapacitors, the switched domains of the target capacitors were relaxed very fast and the size of switched domains rapidly decreased to the half of its initial size within 30 min. Moreover, a higher bias pulse amplitude leads to a much faster back-switching at the target capacitors [Figs. 1(b)(*v*) and 1(b)(*vi*)]. This might originate from a certain charge injection during the switching since higher applied electric fields increase the propensity to a ferroelastic switching of the BFO thin films as well as the amount of injected charge.[12,13] In contrast, the switched domains of the neighbouring BFO capacitors were relatively stable. Comparing the two systems, the switched domains in PZT nanocapacitors were much more stable than in BFO nanocapacitors.

There might be two possible reasons for this cross-talk effect: one might be the different distance between capacitors, and the second might originate in different local structural defects in the ferroelectric film and at the film/electrode interfaces of the capacitor boundary. In order to clarify this, the electric potential and field distributions were simulated using finite element analysis for the different capacitor and switching parameters such as aspect ratio of the system, distance between nanocapacitors, and bias amplitude, by keeping the film thickness constant. Figures 2(a) and 2(b) show the electric potential and field distributions at an applied bias of +2 V under the same scaling dimensions as the present PZT nanocapacitors. For an applied bias of +2 V, the stray field of the target capacitor does not affect the neighbouring capacitors. This case corresponds to the previous reports.[8,9] However, for a higher applied bias of +5 V, the region in which the stray field is present is significantly enlarged. The location $l_C$ at the film surface, measured from the edge of the target capacitor, at which the field becomes equal to the coercive field touches the neighbouring capacitors as shown in Fig. 2(c) by the black arrow. This indicates that for such high applied bias amplitude, domain switching of the neighbouring capacitors can be activated by the stray electric field of the target capacitor.

Figure 3(a) shows $l_C$ of the nanocapacitors with different aspect ratio and capacitor distance under an applied bias of +5 V. As shown in Fig. 3(a), there are three different regions which characterize the cross-talk. If the distance between capacitors is sufficiently large, the electric field distribution and $l_C$ are always the same at each aspect ratio (Region *A*). However, as the capacitor distance decreases, the electric field distribution starts to be distorted from a certain point, which is defined as the *AB* boundary point, due to the influence from neighbouring capacitors (Region *B*). On further decreasing the capacitor distance, the region with a field equal to the coercive field finally touches the neighbouring capacitors and activates domain switching in the neighbouring capacitors (Region *C*). This point is defined as the *BC* boundary point. For an aspect ratio larger than 2, both electric field distribution and $l_C$ are not any more significantly affected by the aspect ratio.[14] This is summarized in Fig. 3 wherein $l_C$ is presented. When the aspect ratio of the system increases, both *AB* and *BC* boundary points increase, and then flatten out. As shown in Fig. 3(b), in order to avoid the cross-talk, the capacitor distance *d* (or the distance/film thickness *d/t* ratio) needs to be larger than the minimum distance defined by the *BC* boundary point. For instance, if the *d/t* ratio is smaller than 1.14 for an aspect ratio of 4, cross-talk can occur under high bias field. Indeed, even if a nanocapacitor has an aspect ratio of larger than 4, the domain propagation towards the neighbouring capacitors could be experimentally observed at a *d/t* ratio of less than 0.5 (not shown here).

More intriguing is the cross-talk generated by the domain wall movement under pulses of different width as well as following the same path by the domain wall movement outside the target capacitor. We might speculate that different local defect states at the capacitor boundary might be responsible for these different effects. We believe that different distributions of local defects can exist along the capacitor boundary and affect domain wall propagation into the neighbouring capacitors. While the results of the finite element analysis show that the domain wall can move into the neighbouring capacitors due to stray fields, the real systems behave differently. This is mostly due to local defects which may hinder the propagation of domain walls into the neighbouring capacitors along particular paths. Indeed, when increasing the bias pulse amplitude as shown in Fig. 1(b), the number of the affected neighbouring capacitors increased.

In summary, we have investigated the switching behaviour and domain wall motion in ferroelectric nanocapacitor arrays under large applied fields and long switching times. Under typical switching pulse conditions, the switched domain is always confined by the capacitor boundary whereas for much longer or higher bias pulses the domain walls start to propagate into the neighbouring capacitors generating a significant cross-talk. The propagation paths to the neighbouring capacitors under repeated runs were always the same and the number of propagated neighbouring capacitors was increased on increasing the bias pulse amplitude. The experimental results on the nanocapacitors and the finite element analysis of the electric field distribution indicate that the observed cross-talk effect can be related to the capacitor parameters combined with local defects in the ferroelectric thin films and at the film/electrode interfaces at the capacitor boundary. The present results can be helpful in understanding ferroelectric domain wall motion at nanoscale with a high relevance to reliability of nanoscale ferroelectric memory devices.

**Acknowledgment**

Y. K. acknowledges the financial support of the Alexander von Humboldt Foundation. We wish to thank A. Schubert and H. You for the helpful comments on the finite element analysis. This work was partly supported by the German Science Foundation (DFG) through SFB762.

Figure 1. PFM phase images of instantaneous domain configurations developing at different (a) bias pulse widths in PZT nanocapacitors and (b) bias pulse amplitudes in BFO nanocapacitors: (a)(*i*) background poled by -4 V for 40 ms, and (*ii-vi*) pulse poled by +4 V for (*ii*) 10 ms and (*iii-vi*) 500 ms. Each case of (*iii-vi*) is 1st, 2nd, 3rd and 4th pulse poling on the same target capacitor. The blue solid lines of (a)(*iii-vi*) indicate the neighbouring capacitors affected by applied long bias pulses. (b) (*i*) background poled by +4 V for 40 ms, and (*ii-vi*) pulse poled by (*ii*) -2 V, (*iii*) -4 V, (*iv*) -6 V, (*v*) -8 V, and (*vi*) -10 V, respectively, for 40 ms. The blue dashed lines of (a)(*i*) and (b) indicate the target capacitor. Scale bars are 50 nm.

Figure 2. (a) Simulated electric potential distribution on nanocapacitors of the same dimensions as the PZT nanocapacitors: a 35 nm thick PZT thin film, a capacitor diameter of 70 nm, a distance $d$ between capacitors of 30 nm, and an applied bias amplitude of +2 V to the central capacitor. (b,c) Simulated electric field distributions under applied bias amplitudes of (b) +2 V and (c) +5 V to the central capacitor. The white region is the region where the electric field is above the coercive field $E_C$ of $4 \times 10^7$ V/m. The black arrow in (c) shows the location $l_C$ of the coercive field at the film surface. Scale bar is 50 nm.

Figure 3. (a) The location $l_C$ at the film surface at which the field becomes equal to the coercive field as a function of the distance $d$ between capacitors with various aspect ratios: 0.1, 0.2, 0.5, 1.0, 2.0, 5.0, and 10.0, respectively. Hollow, filled, and half-filled triangles correspond to the three different regions *A*, *B*, and *C*, respectively. (b) The dependence of the capacitor distance $d$ on the aspect ratio of the system ($t$: film

thickness) for *AB* and *BC* boundary points (see text). The blue solid lines in (b) mark the values above which the aspect ratio does not play any longer a major role.

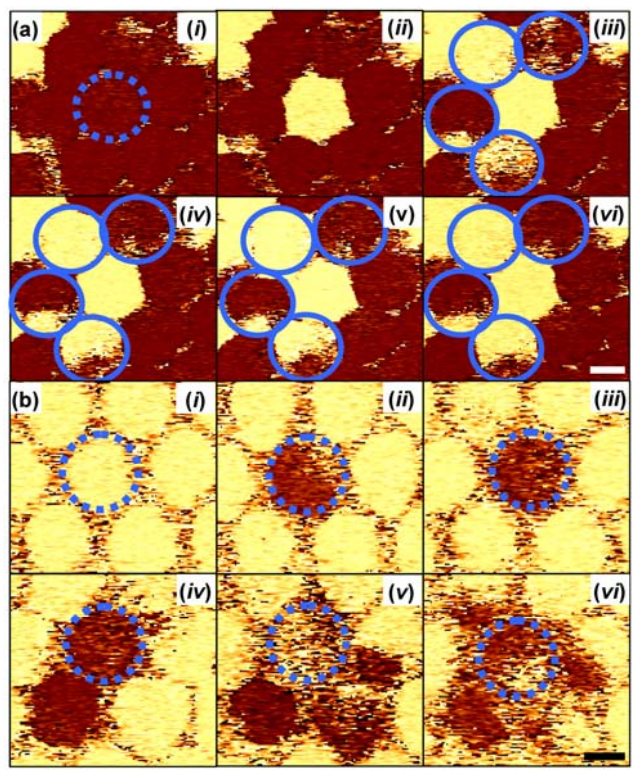

Figure 1. Yunseok Kim et al.

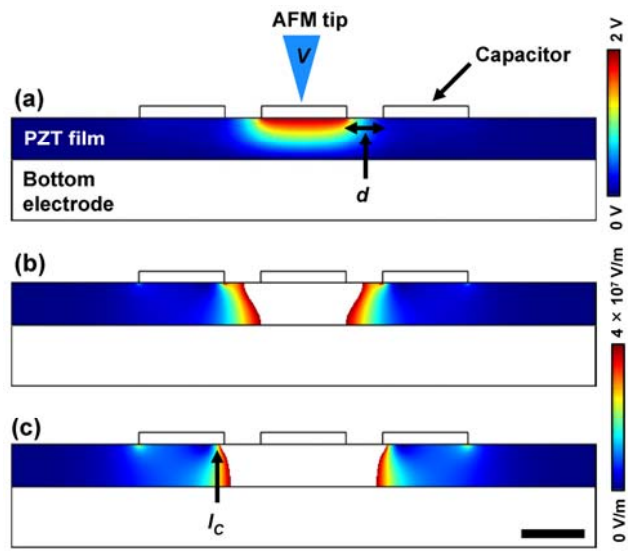

Figure 2. Yunseok Kim et al.

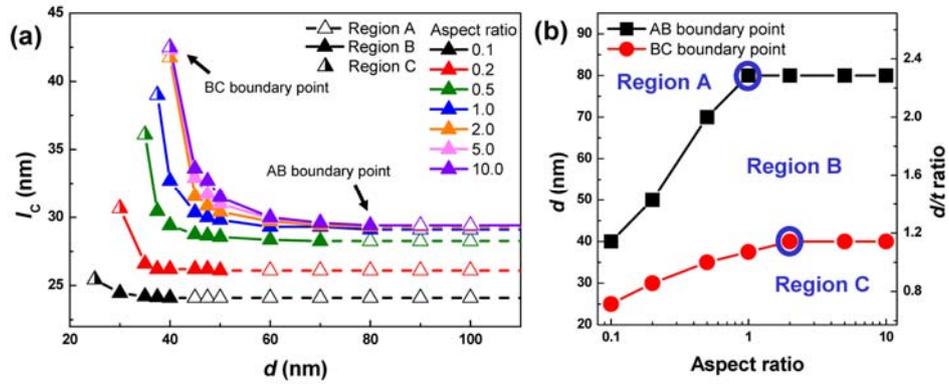

Figure 3. Yunseok Kim et al.

[14] Even though *AB* and *BC* boundary points are not significantly affected by the aspect ratio larger than 2 in Fig. 3(b), there seems to be a slight departure from this rule in 40-50 nm in Fig. 3(a) which is probably related to the different grid size in the model.